\documentclass[%
 aip,
 amsmath,amssymb,
 reprint,%
]{revtex4-1}

\usepackage{graphicx}
\usepackage{dcolumn}
\usepackage{bm}

\usepackage[utf8]{inputenc}
\usepackage[T1]{fontenc}
\usepackage{mathptmx}
\usepackage{etoolbox}

\makeatletter
\def\@email#1#2{%
 \endgroup
 \patchcmd{\titleblock@produce}
  {\frontmatter@RRAPformat}
  {\frontmatter@RRAPformat{\produce@RRAP{*#1\href{mailto:#2}{#2}}}\frontmatter@RRAPformat}
  {}{}
}%
\makeatother
\begin{document}

\preprint{AIP/123-QED}

\title{Strong coupling of plasmonic bright and dark modes with two eigenmodes of a photonic crystal cavity}
\author{Fanqi Meng}%
\email{fmeng@physik.uni-frankfurt.de.}
\affiliation{
Physikalisches Institut, Johann Wolfgang Goethe-Universit\"{a}t, Frankfurt am Main, Germany
}%

\author{Lei Cao}%
\affiliation{ 
Physikalisches Institut, Johann Wolfgang Goethe-Universit\"{a}t, Frankfurt am Main, Germany
}%
\affiliation{ 
State Key Laboratory of Advanced Electromagnetic Engineering and Technology, Huazhong University of Science and Technology, Wuhan 430074, China
}%

\author{Aristeidis Karalis}%
\affiliation{ 
Research Laboratory of Electronics, Massachusetts Institute of Technology, Cambridge,USA
}%

\author{Hantian Gu}%
\affiliation{ 
Physikalisches Institut, Johann Wolfgang Goethe-Universit\"{a}t, Frankfurt am Main, Germany
}%

\author{Mark D. Thomson}%
\affiliation{ 
Physikalisches Institut, Johann Wolfgang Goethe-Universit\"{a}t, Frankfurt am Main, Germany
}%

\author{Hartmut G. Roskos}
\email{roskos@physik.uni-frankfurt.de.}
\affiliation{
Physikalisches Institut, Johann Wolfgang Goethe-Universit\"{a}t, Frankfurt am Main, Germany
}%

\date{\today}

\begin{abstract}
Dark modes represent a class of forbidden transitions or transitions with  weak dipole moments between energy states. Due to their low transition probability, it is difficult to realize their interaction with light, let alone achieve the strong interaction of the modes with the photons in a cavity. However, by mutual coupling with a 
bright mode, the strong interaction of dark modes with photons is possible. This type of mediated interaction is widely investigated in the metamaterials community and is known under the term \textit{electromagnetically induced transparency} (EIT). Here, we report strong coupling between a plasmonic dark mode of an EIT-like metamaterial with the photons of a 1D photonic crystal cavity in the terahertz frequency range. The coupling between the dark mode and the cavity photons is mediated by a plasmonic bright mode, which is proven by the observation of a frequency splitting which depends on the strength of the inductive interaction between the plasmon bright and dark modes of the EIT-like metamaterial. In addition, since the plasmonic dark mode strongly couples with the cavity dark mode, we observes four polariton modes. The frequency splitting by interaction of the four modes (plasmonic bright and dark mode and the two eigenmodes of the photonic cavity) can be reproduced in the framework of a model of four coupled harmonic oscillators. 
\end{abstract}

\maketitle

%

\section{Introduction}
Dark states are either excitations of (quasi-)particles coupled with the ground state by electromagnetic transitions with weak or vanishing dipole moments or electromagnetic modes of resonators which do not or only weakly couple to external radiation. As dark states are not easily accessible by light, it is difficult to achieve strong interaction with the photons of a cavity \cite{Rivera2016}. In recent years, there has been a significant interest to investigate ways to achieve strong coupling of dark states\cite{White2019,Rousseaux2020,Cuartero-Gonzalez2018}. Such investigations are not only of interest for a fundamental understanding of light-matter and many-particle interactions, but also have the potential to provide alternative routes for designing new optoelectronic and memory devices\cite{Zhang2015}. 

EIT-like planar plasmonic metamaterials (MMs) is an intensively studied research topic\cite{Zhang2008,Zhang2016}. It is a classical analogue of EIT of three-level quantum systems, and manifests as a sharp transparency band in a broad absorption spectrum, originating from the destructive interference between a bright and a dark plasmonic mode of the MM. Owing to their highly dispersive refractive index, EIT-like MMs are ideal candidates for realizing slow-light devices \cite{Gu2012,Ma2021} and ultrasensitive sensors\cite{Zhu2021,Wu2017}. The strong coupling between plasmonic bright-bright and bright-dark modes was observed to occur in various MMs structures \cite{Singh2009,Zhang2016,Li2017,Yu2018,Zhang2022}. 

A rotationally symmetric 1D photonic crystal cavity exhibits two degenerate fundamental eigenmodes with perpendicular polarization directions of the electric field. Frequency-resonant, linearly polarized incident photons excite only one of these eigenmodes, which is called the \textit{bright} photonic eigenmode. The unexcited mode with perpendicular polarization can be regarded as \textit{dark}  photonic mode. In previous publications, we reported the strong coupling between bright plasmonic and bright photonic modes\cite{Meng19,Meng2021,Cao22}. In this contribution, we investigate the strong coupling of bright and dark plasmonic modes of an EIT-like MM with bright and dark cavity modes of a terahertz photonic crystal cavity. The interactions among plasmonic bright mode, dark mode and the two photonic eigenmodes can be simulated in a conventional four-particle coupling picture.

\section{Experimental results and electromagnetic simulations}
\begin{figure*}
\includegraphics[width=12cm]{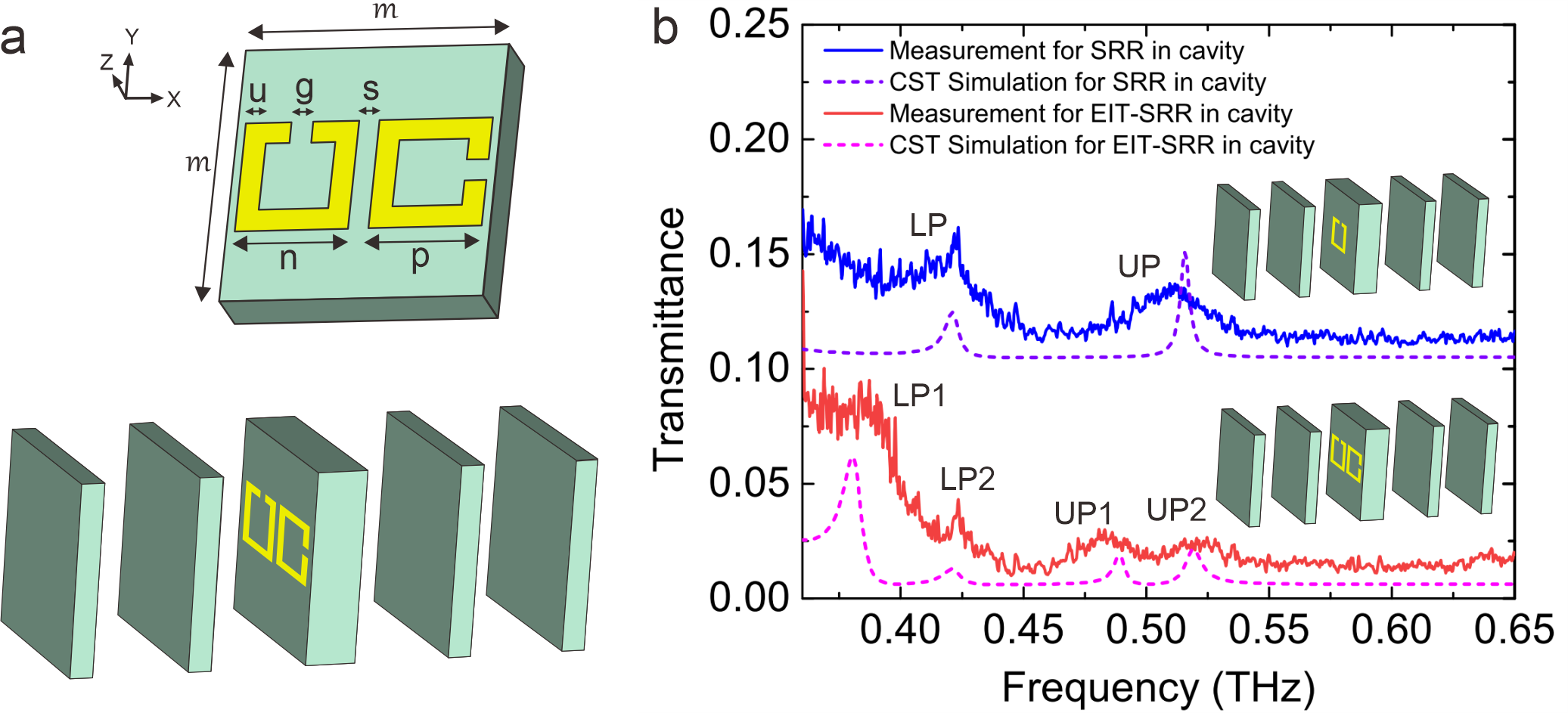}
\caption{\label{fig:exp1}(a) Top: Unit cell of the EIT-like MM; bottom: Schematic presentation of the 1D PC cavity loaded with the EIT-like MM. Geometrical parameters of the EIT-like MM: m =~90~$\mu$m, n = p =~40~$\mu$m, u =~5~$\mu$m, g =~5~$\mu$m, s =~2~$\mu$m. (b) Power transmittance spectra of the metamaterial-loaded cavity, both measurements (full lines) and results of CST simulations (dahed lines). Top panel: MM with a single SRR in the unit cell; bottom pannel: EIT-like MM.}
\end{figure*}

In the experiments, we employed a 1D photonic crystal (PC) cavity made from Si slabs separated by air gaps\cite{Meng19,Zhan16}. 
The slabs, with lateral dimensions of 10$\times$10~mm$^{2}$, were cut from commercially available wafers of highly resistive silicon (specific resistance: >20~k$\Omega$cm, two-side polished, thickness as purchased). For their positioning at a fixed distance, we employed metallic shim rings purchased from MISUMI Europa GmbH, see\cite{Meng2021,Meng19}.
The cavity consisted of five Si slabs. It was composed of a central 100-$\mu$m-thick Si slab as a ‘defect’ layer between two Bragg mirrors. The Bragg mirrors each consisted of two pairs of thin Si dielectric layers and air layers, with the respective thicknesses of 50 $\mu$m and 96 $\mu$m, resulting in a fundamental cavity mode at 490~GHz as confirmed by terahertz transmission measurements. The EIT-like MM consisted of a 2D periodic array of pairs of square-shaped split-ring resonators (SRRs) which were fabricated by photo-lithography on the surface of the defect layer, as shown in Fig.~\ref{fig:exp1}a (top). The SRRs was made by evaporating Ti/Au with the thicknesses of 10/200 nm. The geometrical parameters for the SRRs are given in the caption of the figure. A schematic view of the cavity, loaded with the EIT-like MM structure, is shown in Fig.~\ref{fig:exp1}a (bottom). A description of the experimental setup for the terahertz measurements can be found in Refs.~\cite{Meng19,Meng2021}. The electric radiation field was polarized in the x-direction (for the definition of directions, see the coordinate system in Fig.~\ref{fig:exp1}a (top)).

For comparative measurements, we used a MM with a single square-shaped SRR in the unit cell. It had the same geometrical period as the SRRs of the EIT-like MM unit cell, the side with the split was oriented along the x-direction.
The resonance frequency as predicted by simulations using the CST electromagnetic solver (CST: \textit{Computer Simulation Technology} from Dassault Systèmes Simulia SE) is 445~GHz. The measured resonance frequency of the bare cavity is 480~GHz, which is slightly lower than the simulated frequency due to fabrication tolerances. With the bright SRR MM in the PC cavity, we measured the transmission spectrum 
shown by the blue solid curve in Fig.~\ref{fig:exp1}b. One observes two transmittance resonances which represent the lower polariton (LP) and the upper polariton (UP) of the coupled system. When the dark SRR MM is positioned in the PC cavity, the measured transmission spectrum only shows one cavity mode, with the reduced Q-factor of ~23. The measured cavity frequency decreases to 466 GHz due to additional metal added to the cavity. By taking the perturbed cavity mode frequency (466 GHz) and simulated resonant frequency of SRR, we determined the Rabi splitting is of 87~GHz, which corresponds to about 19\% of the cavity resonance frequency and hence indicates a mode interaction in the strong-coupling regime \cite{Meng19}. Measured transmittance peaks and resonance frequencies (dashed line) obtained by CST simulations are in good agreement with each other. But the measured spectral linwidths are broader than that of the simulations. 

When instead the EIT-like MM is used in the cavity, one instead observes four polariton modes. This is shown in the lower panel of  Fig.~\ref{fig:exp1}b (red solid curve). 
The four polariton modes are at 387~GHz (LP1), 423~GHz (LP2), 483~GHz (UP1) and 520~GHz (UP2), in good agreement with CST simulations (dashed line in Fig. 1b). The appearance of four polariton modes can be explained as follows.
 
 \begin{figure}
\centering\includegraphics[width=8cm]{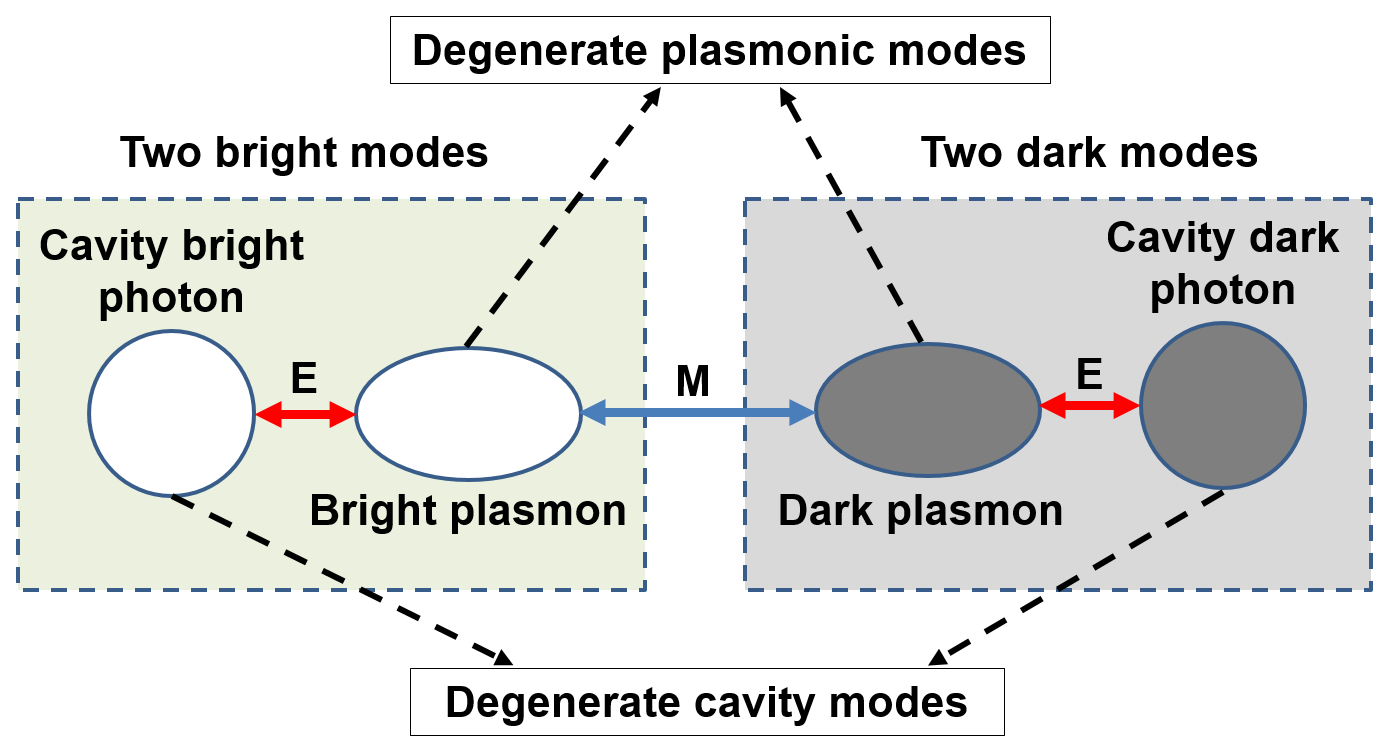}
\caption{\label{fig:model1}Schematic illustration of the four-mode coupling. "E" represents electric coupling and "M" represents magnetic coupling.}
\end{figure}
 
The mutual interaction of the modes of the cavity and the EIT-like MM is schematically shown in Fig.~\ref{fig:model1}. One can understand the interaction if one applies a sequential-excitation picture: For an incoming terahertz wave with an electric field polarized along the x-axis, the photons excite the bright cavity eigenmode, which couples directly with the plasmonic bright mode of the left SRR in Fig.~\ref{fig:exp1}a. The right SRR (exhibiting the plasmonic dark mode, frequency-degenerate with the bright mode of the left SRR) is then 
excited mainly due to magnetic near-field interaction -- albeit also with an electric contribution \cite{Li2017} -- with the left SRR, with a coupling strength dependent on the lateral distance between them. The right SRR becomes radiative, with its electric dipole moment oriented along y-direction contributing emission with an electric field in y-direction which couples to the dark cavity eigenmode. 
This interaction chain hence couples all four quasi-particles (plasmonic bright and dark MM modes, photonic bright and dark cavity modes) with each other, which produces four new polariton modes as well as cross-polarization conversion.
 
\begin{figure*}
\centering\includegraphics[width=12cm]{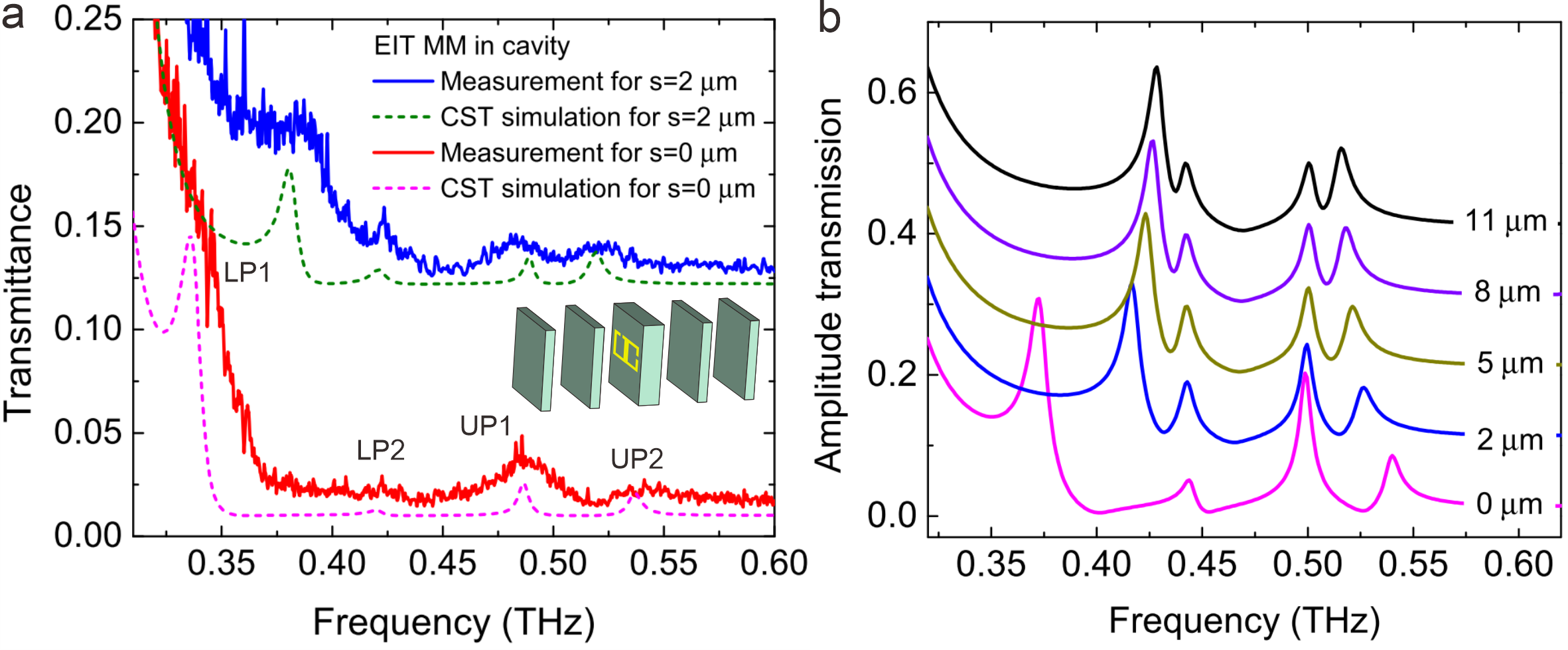}
\caption{\label{fig:exp2} (a) Power transmission spectra of radiation polarized in x-direction. For a cavity loaded with an EIT-like MM, where the SRRs are physically connected with each other (experiment: red line, CST simulation: green dashed line), respectively have a distance s of 2~$\mu$m (experiment: blue line, CST simulation: magenta dashed line). The 2-$\mu$m-data are a reproduction of those shown already in Fig.~\ref{fig:exp1}b.
(b) Amplitude transmission spectra of radiation polarized in x-direction. CST simulations for several distances between the two SRRs. Each neighboring curve shifts vertically by 0.1 for better comparison. In this set of simulations, m is taken as 110~$\mu$m, and n and p are assumed to be 38~$\mu$m. 
}
\end{figure*}

As stated above, the Rabi splitting between polariton modes LP1 (UP1) and LP2 (UP2) depends on the distance of the two SRRs. This is confirmed both experimentally and theoretically with the data shown in Fig.~\ref{fig:exp2}. The red solid curve in Fig.~\ref{fig:exp2}a displays the four polariton modes measured when the distance between the two SRRs is zero and the two SRRs are physically connected with each other. The LP1 mode lies in the shoulder of the PC's forbidden band, but is still recognizable. For comparison, the blue curve in Fig.~\ref{fig:exp2}a reproduces the transmittance curve of Fig.~\ref{fig:exp1}a measured with the EIT-like MM in the cavity, but here with the SRRs separated by 2~$\mu$m. In the case of zero distance, the mode splitting between LP1 and LP2, respectively UP1 and UP2, is larger than if the distance of the two SRRs is 2~$\mu$m. This observation is confirmed by the results of CST simulations shown by the dashed curves in Fig.~\ref{fig:exp2}a. When the two SRRs are not physically connected, the coupling between them is mediated by the light-induced magnetic field, and thus is \textit{inductive}. 
When the two SRRs are physically connected, the coupling between them is both \textit{inductive} and \textit{conductive}, which leads to a stronger interaction and larger Rabi splitting \cite{Liu2008}. 

Additional insight is provided by CST simulations. Fig.~\ref{fig:exp2}b shows calculated amplitude transmission spectra for various distances between the two SRRs, ranging from 0~$\mu$m to 11~$\mu$m. In order to suppress interaction between them for all separation distances, the period of the MM was increased from 90~$\mu$m in the measurements to 110~$\mu$m in the simulations, and the size of the SRRs was reduced from 40~$\mu$m to 38~$\mu$m. The polariton splitting is then always dominated by interactions within each unit cell. The simulation confirms that the splitting between LP1 and LP2, and UP1 and UP2, increases as the distance between the SRRs is reduced. Interestingly, the frequency positions of LP2 and UP1 do not change much with distance. Except for zero distance, LP2 remains at 420~GHz and UP1 at 500~GHz. As the distance between the two SRRs increases, LP1 moves upwards in frequency towards LP2, and UP2 downwards towards UP1. One anticipates that ultimately the two LP modes would merge into one, and the same would occur with the two UP modes, hence the four polariton modes would converge into the two polariton modes which arise by the coupling of the bright SRR mode with the bright cavity mode.

\begin{figure}
\centering\includegraphics[width=8cm]{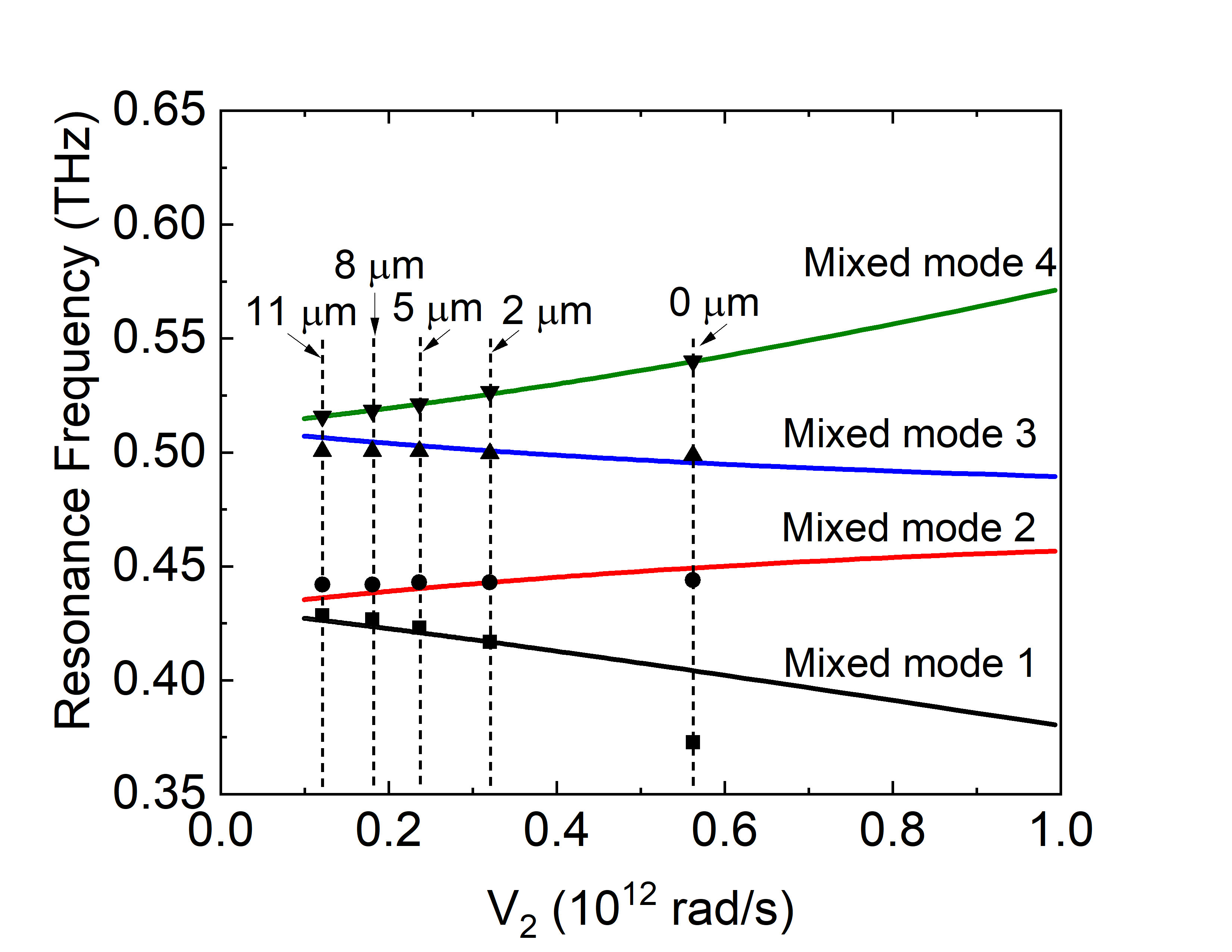}
\caption{\label{fig:model10}Resonance frequencies of the four mixed modes (representing the polaritons) as obtained by the model of four coupled harmonic oscillators. Black symbols: Resonance frequencies are extracted from Fig.~\ref{fig:exp2}b, and plotted versus the value of the coupling parameter $V_2$ obtained by a fit. The full lines are the results of a fit with a fixed global coupling parameter $V_1$.  
}
\end{figure}

\section{Discussions}
\subsection{Model of four coupled oscillators}
\label{FCO}
The frequency dependence of the interaction among the four modes can be captured with a model of four coupled harmonic oscillators, described by four differential equations of motion with two coupling constants, $V_1$ and $V_2$. $V_1$ represents the coupling constant between plasmon and photon, and $V_2$ indicates the coupling constant between bright plasmon and dark plasmon. The set of differential equations together with the derivation of the eigenfrequenices of the mixed modes is given in Appendix \ref{Append_EoM}, extending the analysis of \cite{Meng2021}.

Each mode is represented by its time-dependent generalized coordinate, $x_1(t)$ to $x_4(t)$. $x_1$ and $x_4$ describe the bright and dark cavity modes, respectively, assumed to be frequency-degenerate, with the common resonance frequency $\omega_c = 2\pi f_c$. $x_2$ and $x_3$ are the coordinates of the bright and the dark plasmonic mode of the MM -- also assumed to be frequency-degenerate at $\omega_p = 2\pi f_p$. The two bright modes couple directly with each other, which is expressed by coupling terms in the equations for $x_1$ and $x_2$ with the coupling strength $V_1$. Correspondingly, the dark modes couple with each other. The interaction is via the y-oriented electric dipole moment of the dark plasmonic mode, which has the same dipole strength as the x-oriented dipole moment of the bright plasmonic mode. In the harmonic-oscillator model, this interaction finds its expression in the equations of motion for $x_3$ and $x_4$ by coupling terms, which also have the strength $V_1$. The third and final interaction is between the two plasmonic modes. It is represented by coupling terms in the equations of $x_2$ and $x_3$, with the strength of this interaction given by $V_2$.  A higher value of $V_2$ corresponds physically to a smaller distance between the two SRRs of the unit cell.

The full lines in Fig.~\ref{fig:model10} show the evolution of the resonance frequencies of the mixed modes (representing the polaritons), obtained from the set of differential equations, as a function of the coupling parameter $V_2$. The curves were derived by a fit to the resonance frequencies extracted from the five CST-determined transmission spectra displayed in Fig.~\ref{fig:exp2}b. The coupling parameter $V_1$ was used as a global parameter for all values of the distance s of the SRRs, with a resultant value of $V_1 = 0.4950\times 10^{12}$~rad/s. The other parameters in the model were obtained as $f_p$ = 464.8~GHz and $f_c$ = 474.2~GHz. 
The mixed-mode branches 1 and 4 are found to exhibit strong gradients as a function of $V_2$, while the branches 2 and 3 are less sensitive to the variation of $V_2$. For $V_2 \rightarrow 0$, branch 1 merges with branch 2, as does branch 3 with branch 4.

We also performed independent fits for each value of the distance s between the SRRs, where only the coupling parameter $V_2$ was varied. The black symbols in Fig.~\ref{fig:model10} represent the resonance frequencies as a function of $V_2$.
The fit with a fixed value of $V_1$ especially predicts a change of the resonance frequencies of the mixed modes 2 and 3, when the value s is changed, while one observes in Fig.~\ref{fig:exp2}b (and for the black circles and upwards-pointing triangles in Fig.~\ref{fig:model10}) that the mixed-mode resonance frequencies remain nearly constant, except for $s \rightarrow 0~\mu$m. One also notes that for $s \rightarrow 0~\mu$m, a fit with a constant $V_1$ overestimates the resonance frequency of the mixed mode 1, and also slightly that of mixed mode 2. This is caused by the conductive coupling which arises in addition to the inductive coupling of the two SRRs and impacts also $V_1$.

\begin{figure}
\centering\includegraphics[width=8cm]{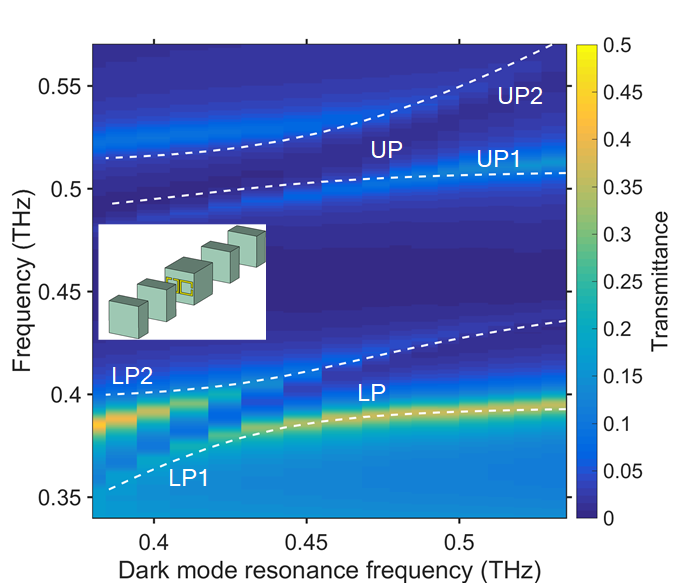}
\caption{\label{fig:darkDisp} Color plots of the simulated transmission spectra of a EIT-MM-loaded cavity as a function of the resonance frequency of the dark SRR. The inset displays the loaded cavity, with the EIT-like MM represented by a two-SRR unit cell. The white dashed lines represent the results calculated with the equations of motion of the four-coupled-oscillators model.}
\end{figure}

The plasmonic dark mode plays a key role in bringing the cavity dark mode into the coupling scenario. We study now how this mediating effect is influenced when the cavity dark mode is detuned from the resonance frequencies of the cavity modes and the plasmonic bright mode.
We simulated the transmission of the EIT-SRR-loaded cavity with the CST Maxwell solver, 
changing the geometrical dimensions of the SRR located at the right side in the sketch of Fig.~\ref{fig:exp1}a (top). For the MM, we otherwise used the same geometrical parameter values as for the plots in Fig.~\ref{fig:exp2}b, namely m = 110~$\mu$m, n = 38~$\mu$m, and s = 2~$\mu$m. Only the value of p was changed from 32~$\mu$m to 42~$\mu$m.
The color graph in Fig.~\ref{fig:darkDisp} presents the transmission spectra for x-polarized radiation as a function of the changing resonance frequency of that SRR.
When the frequency is strongly detuned (both at the low-frequency limit and the high-frequency one), both the plasmonic dark mode and the cavity dark mode are no longer involved in the strong coupling, and only the bright modes of the cavity and the MM participate in the strong interaction. One thus expects only two polariton modes to remain, which lie around the frequencies marked as ``LP'' and ``UP'' (written in white lettering) in Fig.~\ref{fig:darkDisp}. In fact, for detuning to large frequencies, the UP2 and the LP2 mixed-mode branches die out, leaving oscillator strength only in UP1 and LP1. For detuning to low frequencies, the opposite occurs: Oscillator strength survives in UP2 and LP2, while UP1 and LP1 die out.
Inbetween these extremes, one observes anti-crossing behavior of the UP1 and UP2 branches, and the LP1 and LP2 branches, the respective 
mode splittings amount to about 8~GHz, are larger than the spectral linewidths of the polariton modes, and correspond to 6\% and 5\% of the respective LP and UP mode frequencies. This indicates that the interactions are in the strong-coupling regime. 

We also calculated the mixed-mode frequencies with the equations of motion of the four coupled harmonic oscillators, using the values $f_c = 462$~GHz, $f_p = 436$~GHz for the resonance of the unmodified SRR, $V_1 = 0.7024 \times 10^{12}$~rad/s and $V_2 = 0.2903 \times 10^{12}$~rad/s.
The results are plotted as white dashed curves in Fig.~\ref{fig:darkDisp}. They reproduce the resonances obtained with the full-wave simulations reasonably well. The deviations are probably due to the fact that we used fixed values for the coupling constants $V_1$ and $V_2$.

We end this chapter with two remarks and refer the reader for further information to the Appendix. First, we point out the equivalent-circuit model addressed in Appendix \ref{Append_Ecm}. The model is based on lumped elements (capacitors, inductors and resistors) from which the various coupled resonators are built.  The model can provide a qualitative and even quantitative analysis of the system of coupled resonators if the coupling mechanisms (capacitive, inductive or both) are clear\cite{Poo2014}. In this regard, it is of a more microscopic nature than the  phenomenological model of four coupled equations of motion. 

Second, we point out that if both SRRs are bright (i.e., have parallel electric dipole moments), there will be three polariton modes instead of four, when coupled to the cavity (see Appendix \ref{Append_bright} with Fig.~\ref{fig:bright}). In this case, the MM modes only interact with the bright cavity mode and not with the dark one. The mixed-mode frequencies are well reproduced by the equations-of-motion model of three coupled oscillators\cite{Meng2021}.

\begin{figure}
\centering\includegraphics[width=8cm]{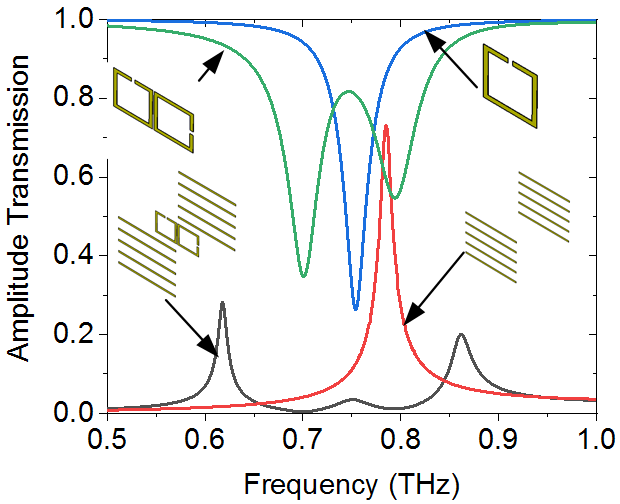}
\caption{\label{fig:FP} Amplitude transmission spectrum of a metallic-grating-type Fabry-Perot cavity loaded with an EIT-like MM (dark-grey line). Amplitude transmission spectrum of empty Fabry-Perot cavity (red line), of bare MM with single bright SRR in each unit cell (blue line), of bare MM with EIT SRRs in each units cell (green line). The sketches indicate the four structures of which the transmission spectra are displayed, which were calculated with CST for radiation polarized perpendicular to the metal wires of the grating.
}
\end{figure}

\subsection{Coupling in a metallic grating-type Fabry-Perot cavity}
A way to suppress the coupling to the cavity dark mode is to employ a cavity that does not support modes with a polarization orthogonal to that of the bright mode.
Such a type of cavity is a Fabry-Perot resonator made with metallic gratings as reflectors. Ohmic losses strongly suppress standing waves with a polarization parallel to the metal stripes of the grating, but the losses are much lower for radiation polarized perpendicular to the stripes. (We note in passing that a fundamental-mode cavity made with reflectors consisting of unpatterned sheets of gold, or silver or any other good metallic conductor would have a poor Q-factor for all polarization directions. In a fundamental mode cavity, the radiation field penetrates into the metal where it experiences substantial losses even in the case of good conductors. For that reason, we employ cavities made from dielectric materials for studies like those of this publication.)

We investigated mode coupling of SRR MMs in a grating-type cavity by CST simulations. The gratings were assumed to be made from rectangular gold wires (conductivity of $4.56\times 10^{7}$ S/m) with a width of 1.4 $\mu$m and a thickness of 0.2 $\mu$m, and separated from each other by 0.6 $\mu$m. The wires were assumed to be free-standing in air. 
The cavity consisted simply of two grating reflectors, with a distance of 180~$\mu$m between them. The red line in Fig.~\ref{fig:FP} displays the transmission spectrum of the empty cavity for the amplitude of radiation polarized perpendicular to the grating wires. The transmittance exhibits a peak at 784~GHz, the resonance frequency of the fundamental cavity mode. 

Into the cavity, centered between the reflectors, we placed an EIT-like MM structure (see schematic drawings in Fig.~\ref{fig:FP}). It consisted of two substrate-free metallic SRRs, each having a fundamental resonance at 754~GHz (see blue line in Fig.~\ref{fig:FP}), split into two polariton modes at 701~GHz and 794~GHz (see green line in Fig.~\ref{fig:FP}). The parameters of the MM in the simulations (using the nomenclature of Fig.~\ref{fig:exp1}a (bottom)) were as follows. Material: gold stripes with a thickness of 0.2~$\mu$m, u = 5 ~$\mu$m, g = 5~$\mu$m, 
s = 2~$\mu$m, n = p = 40~$\mu$m, m = 90~$\mu$m. With the EIT-like MM structure in the cavity, one observes three transmission peaks at 618~GHz, 751~GHz and 862~GHz, the middle one lying close to the resonance frequency of the individual SRRs and being considerably weaker than the transmission peaks above and below. The appearance of three resonances is consistent with the strong coupling among three modes (cavity bright mode, plasmonic bright mode and plasmonic dark mode). The suppressed cavity dark mode does not contribute noticeably in the interaction. This result supports the notion that it is indeed the cavity dark mode which is responsible for the appearance of the fourth resonance in the experiments and simulations with the EIT-like MM in the dielectric cavity discussed in the previous sections of this publication.  

\subsection{Cross-polarization conversion and polarization tuning}
\begin{figure*}
\centering\includegraphics[width=12cm]{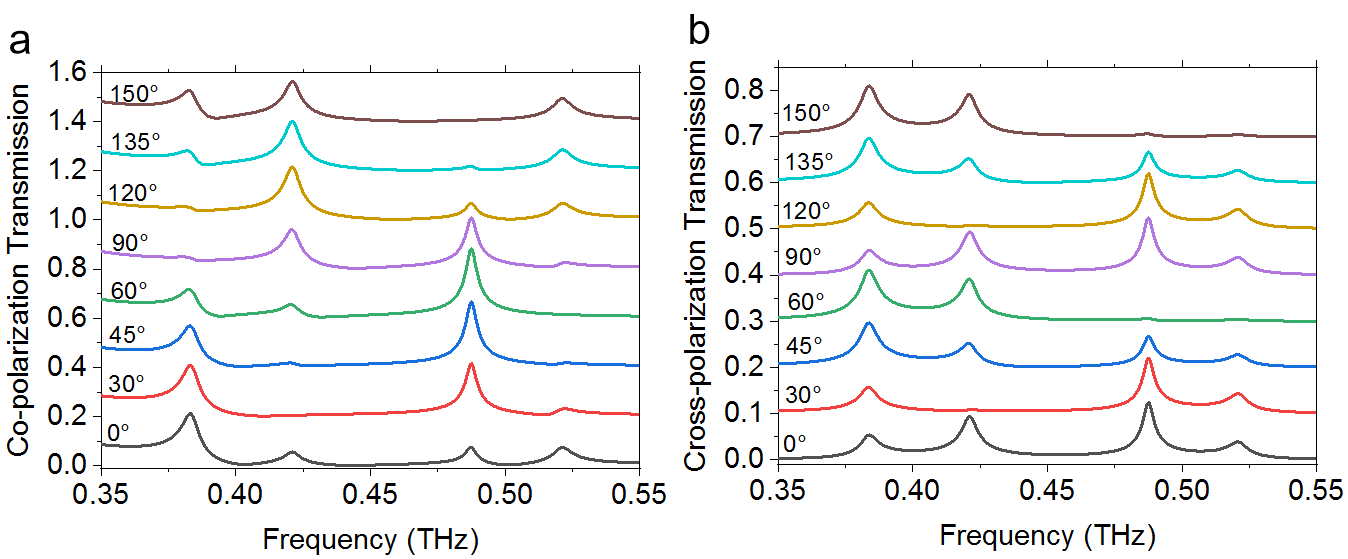}
\caption{\label{fig:tunable} Amplitude transmission spectra of linearly polarized radiation impinging onto the EIT-like-MM-loaded cavity as a function of the polarization direction of the incident electric field (CST simulations). 
(a) Co-polarized transmitted field component, the spectra for different angles are displaced vertically by 0.2. (b) cross-polarized transmitted field. For clarity, the spectra for different angles are displaced vertically by 0.1.} 
\end{figure*}

The coupling of the bright and dark modes of the EIT-like MM can be employed for cross-polarization conversion. Linearly polarized incident radiation is converted 
to elliptically polarized one\cite{Li2017}. 
We briefly consider here, how the orientation of the incoming electric field influences the polarization of the outgoing field. Fig.~\ref{fig:tunable} displays the co-polarized and cross-polarized transmission spectra of the radiation passing the EIT-like-MM-loaded cavity, as the polarization angle $\phi$ of the incident electric field is varied from 0$^{\circ}$ to 150$^{\circ}$. Co-polarization (cross-polarization) means that the field orientation of the outgoing wave parallel (perpendicular) to the linear field orientation of the incoming wave is displayed. We do not analyze the phases of the two outgoing waves relative to the incoming one. 
The structural parameters of the EIT-like MM and the cavity are the same as those used for Fig.~\ref{fig:exp1}b, and an orientation of $\phi =0^{\circ}$ corresponds to the situation shown there. 

In Fig.~\ref{fig:tunable}, for most values of $\phi$ both the co- and cross-polarized transmission spectra exhibit four resonance peaks, and their frequencies do not change with the variation of $\phi$, only the amplitudes. The co-polarized transmission exhibits a $180^{\circ}$ rotational symmetry, the cross-polarized transmission, however, a $90^{\circ}$ symmetry.
In co-polarized transmission, one observes at specific angles only three polariton modes. If we number the modes with rising frequency as $\#1$ to $\#4$, then within the range $0^{\circ} \leq \phi < 180^{\circ}$, mode $\#2$ disappears at about $30^{\circ}$, mode $\#4$ at about $60^{\circ}$, mode $\#1$ at about $100^{\circ}$, and mode $\#3$ at about $150^{\circ}$. A similar reduction of the number of observed modes occurs at specific angles in cross-polarized transmission. The details of these phenomenological findings are intriguing and invite future studies. 


\section{Conclusion}
In conclusion, we have studied experimentally and theoretically the transmission properties of a metamaterial, consisting of coupled split-ring resonators and exhibiting electromagnetically-induced-transparency-like behavior, in a rotationally symmetric photonic crystal cavity at sub-terahertz frequencies. We have observed the appearance of four transmission peaks. They have been identified as the signatures of polariton modes arising from strong mutual coupling of four modes, the bright and dark fundamental modes of the metamaterial with the two orthogonally polarized fundamental cavity modes. In a quasi-particle picture, the mutual coupling occurs between bright and dark plasmons of the metamaterial and the orthogonally polarized photons of the cavity. Incoming linearly polarized radiation is converted to elliptically polarized radiation. The coupling to both cavity modes can be reduced to coupling to only one (and the appearance of only three instead of four polariton modes), if either the rotational symmetry of the cavity is destroyed (e.g. by using a cavity made from metallic-grating reflectors) or if the metamaterial exhibits a total dipole moment which is parallel to the polarization of the incoming wave. 


\section*{Acknowledgements}
This research work was funded by DFG projects RO 770/46-1 and RO 770/50-1 (the latter being part of the  DFG-Schwerpunkt "Integrierte Terahertz-Systeme mit neuartiger Funktionalität (INTEREST)"). Lei Cao acknowledges support from the HUST Overseas Training Program for Outstanding Young Teachers.

\section*{AUTHOR DECLARATIONS }
\subsection*{Conflict of interest}
The authors have no conflicts to disclose.
\subsection*{Author contributions}
\textbf{Fanqi Meng:}Conceptualization (lead); Investigation (lead); Writing-original draft (equal); Formal analysis (equal).
\textbf{Lei Cao:} Writing-original draft (equal); Formal analysis (equal); Validation (equal).
\textbf{Aristeidis Karalis:} Validation (equal).
\textbf{Hantian Gu:} Investigation (equal).
\textbf{Mark D. Thomson:} Conceptualization (equal).
\textbf{Hartmut G. Roskos:} Conceptualization (equal); Writing-review \& editing(equal); Funding acquisition (lead); Supervision (lead).

\section*{Data Availability }
The data that support the findings of this study are available from the corresponding author upon reasonable request.

\appendix

\section{Coupled differential equations of the four-oscillators model}

\begin{verbatim}

\end{verbatim}

\label{Append_EoM}
We assume four harmonic oscillators, two having a resonance frequency $\omega_1$ = $\omega_4$ = $\omega_c$ (cavity mode frequency) and the other two being resonant at $\omega_2$ = $\omega_3$ = $\omega_p$ (MM plasmonic mode frequency). Oscillators 1 and 2, respectively 3 and 4, are coupled, with the coupling strength being determined by the parameter $V_1$. Also, oscillators 2 and 3 are coupled, with a strength given by $V_2$. For an explanation of this choice of coupling, see Sec.~\ref{FCO}. The differential equations of motion for the four generalized coordinates $x_1(t)$ to $x_4(t)$ of the oscillators read as follows:
\begin{equation}
\begin{aligned}
    & \ddot{x}_1 + \omega_c^2x_1 + V_1\dot{x}_2 = 0\,,  \\
    & \ddot{x}_2 + \omega_p^2x_2 - V_1\dot{x}_1 - V_2\dot{x}_3 = 0 \,,  \\
    & \ddot{x}_3 + \omega_p^2x_3 + V_2\dot{x}_2 + V_1\dot{x}_4 = 0 \,, \\
    & \ddot{x}_4 + \omega_c^2x_4 - V_1\dot{x}_3 = 0 \,.
\end{aligned}
\end{equation}

If the time-dependent factor is taken as $x_i(t)$ = $x_i^0e^{-j{\omega}t}$, with $\,i = 1, 2, 3, 4$, the frequency-domain expressions of the differential equations can be written in matrix form as
\begin{equation}
    \begin{bmatrix}
    {\omega}^2-{\omega}_c^2 & j{\omega}V_1 & 0 & 0  \\
    -j{\omega}V_1 & {\omega}^2-{\omega}_p^2 & -j{\omega}V_2 & 0  \\
    0 & j{\omega}V_2 & {\omega}^2-{\omega}_p^2 & j{\omega}V_3  \\
    0 & 0 & -j{\omega}V_3 & {\omega}^2-{\omega}_c^2
    \end{bmatrix}
    \begin{bmatrix}
     \;  x_1^0 \; \\
      x_2^0 \\
      x_3^0 \\
      x_4^0 
    \end{bmatrix}
    = 0 \,.
\end{equation}

Letting the determinant be zero, we obtain the function for the eigenfrequencies in general polynomial form:
\begin{equation}
    ax^4 + bx^3 + cx^2 + dx + e = 0 \,,
\end{equation}
with $x= \omega^2$. The coefficients are given explicitly by
\begin{equation}
\left\{
\begin{aligned}
& a = 1 \,, \\
& b = -(2{\omega}_c^2+2{\omega}_p^2+2V_1^2+V_2^2) \,, \\
& c = 2{\omega}_c^2{\omega}_p^2+({\omega}_c^2+{\omega}_p^2)({\omega}_c^2+{\omega}_p^2+2V_1^2)+2{\omega}_c^2V_2^2+V_1^4  \,, \\
& d = -[2{\omega}_p^2{\omega}_c^2({\omega}_p^2+{\omega}_c^2)+2{\omega}_p^2{\omega}_c^2V_1^2+{\omega}_c^4V_2^2] \,, \\
& e = {\omega}_c^4{\omega}_p^4 \,.
\end{aligned}
\right.
\end{equation}

Numerical calculation is used to solve the equation and obtain the four eigenfrequencies of the coupled harmonic oscillators.

\section{Equivalent-circuit model}
\label{Append_Ecm}
\begin{figure*}
\centering\includegraphics[width=12cm]{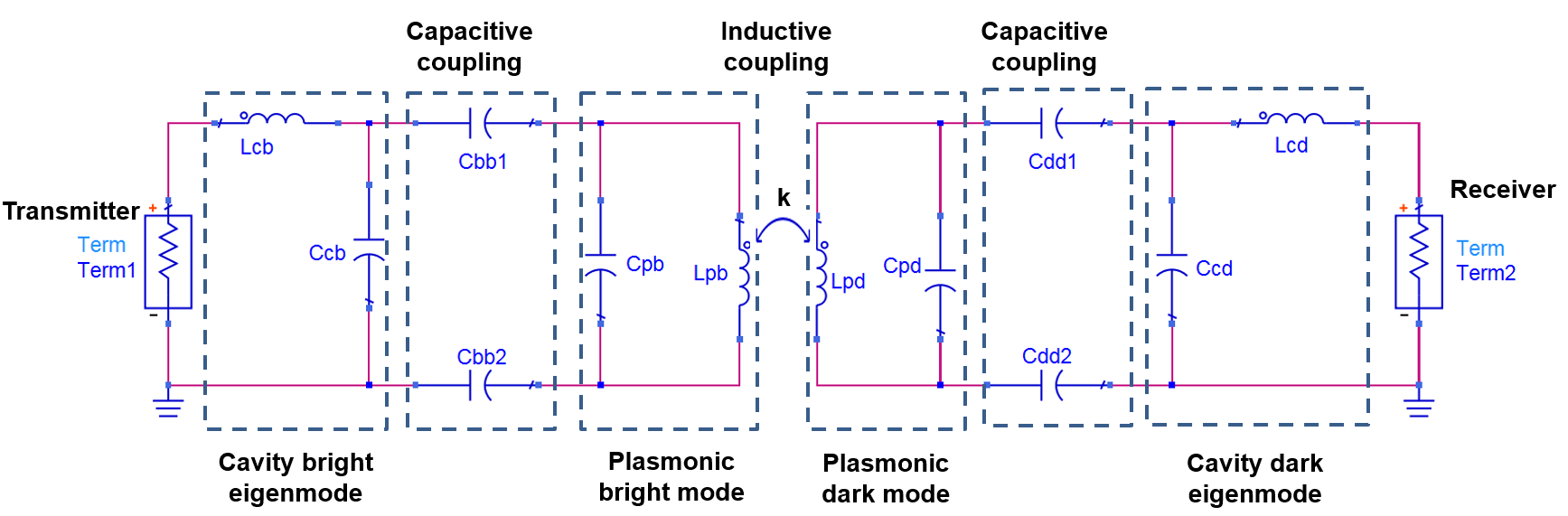}
\caption{\label{fig:model2}Equivalent-circuit model for the four-mode coupling phenomenon in the photonic cavity. 
}
\end{figure*}

An alternative way to physically simulate the interaction of the photonic and plasmonic resonators is to represent them by LC oscillators of an electronic circuit whose wiring expresses the interaction between the oscillators \cite{Poo2014}. If damping is to be included, one could employ LCR oscillators, but we will only considered the undamped limit here. Fig.~\ref{fig:model2} shows an equivalent circuit which implements the model of Fig.~\ref{fig:model1} for the description of the four-mode coupling phenomenon in the PC cavity. Four LC sub-circuits are used to model the four resonators. The coupling between the cavity photonic modes and the plasmonic MM modes is assumed to be capacitive (mediated by the electric radiation field), while the coupling between the plasmonic bright and dark modes is considered as inductive (mediated by the magnetic field of the MM). 

With the formula for approximate inductance calculations of Ref.~\onlinecite{Sydoruk2009,Zhu2014}, we obtain for each SRR, with the dimensions as given in Fig.~\ref{fig:exp1}, an equivalent inductance of Lpb = Lpd = 0.14~nH. The equivalent inductance of the photonic cavity is chosen arbitrarily as Lcb = Lcd = 1~nH. The equivalent capacitances of the SRRs are determined as Cpb = Cpd = 0.91~fF to yield a resonance frequency of the SRRs of $f_p= 445$~GHz, similarly the equivalent capacitances of the cavity are chosen to be Ccb = Ccd = 0.11~fF to yield a cavity resonance at $f_c= 480$~GHz.
For the coupling capacitance, we obtain Cbb1 = Cbb2 = Cdd1 = Cdd2 = 0.09~fF so that the resonance frequencies obtained from the circuit model match closely with those obtained from CST simulation, while the dimensionless inductive coupling coefficient k between Lcb and Lcd can be varied from 0 to 1, depending on the lateral distance of the two SRRs. The losses in the system are neglected. 

\begin{figure*}
\centering\includegraphics[width=12cm]{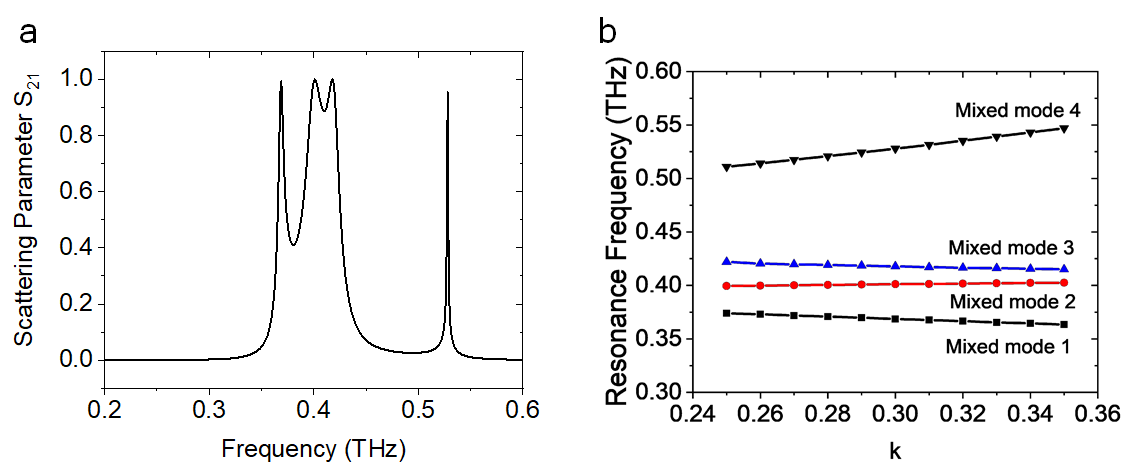}
\caption{\label{fig:model3}(a) Calculated transmission coefficient of the circuit shown in the preceding Fig.~\ref{fig:model2}, for an inductive coupling coefficient k = 0.3. (b) Frequencies of the four eigenmodes as a function of the inductive coupling coefficient k, with $0.25\leq \mathrm{k} \leq 0.35$. 
The port impedance was chosen to be 100~$\Omega$.}
\end{figure*}

Fig.~\ref{fig:model3}a shows the calculated scattering parameter S$_{21}$ for the value k = 0.3 of the inductive coupling coefficient. One observes four distinct resonances corresponding to the four eigenmodes. If the coupling coefficient is varied between 0.25 and 0.35, one obtains the evolution of the four resonance frequencies as shown in Fig.~\ref{fig:model3}b. With the increase of k, corresponding to a decrease of the lateral distance between the two SRRs, the highest branch (mixed mode 4) goes up in frequency, while the lowest branch (mixed mode 1) moves down. The middle two branches remain almost unchanged. These results reproduce the general trends of Figs.~\ref{fig:exp2}b and \ref{fig:model10}, which suggests that the simple equivalent circuit can serve as a tool to model the four-mode coupling process. The predictive power of the model is, however limited to semi-quantitative statements, because the exact values of the eigenfrequencies cannot be predicted due to the uncertainty of the values of many circuit parameters and the omission of additional capacitive coupling between the two SRRs\cite{Poo2014}.The precise determination of element parameters in the circuit model is beyond the scope of this publication.

\section{The case of two bright SRRs in the 1D PC cavity}
\label{Append_bright}
\begin{figure*}
\centering\includegraphics[width=12cm]{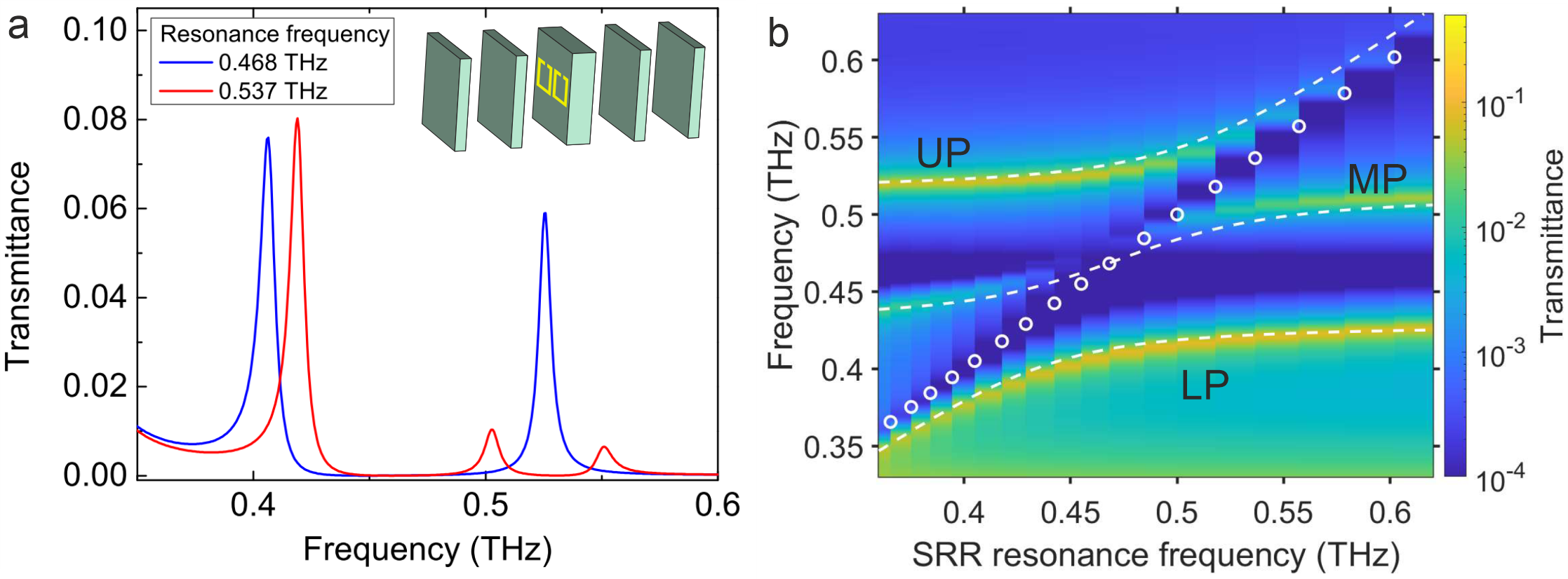}
\caption{\label{fig:bright}(a) the simulated co-polarized transmittance spectra of two bright SRRs loaded 1D PC cavity. The resonance frequency of the cavity lies at 480~GHz, and the resonance frequency of one SRR is fixed at 468~GHz. The resonance frequencies of another SRR lie at 468~GHz (blue curve) and 537~GHz (red curve). The inset displays the SRRs-loaded cavity, with the SRRs represented by a two-SRR unit structure. (b) Color plots of the simulated co-polarized transmission spectra of two bright SRRs loaded 1D PC cavity, as a function of the resonance frequency of one SRR. In these simulations, m is taken as 110~$\mu$m, n is assumed as 38~$\mu$m. The resonance frequencies of the cavity and one SRR are fixed at 0.468 and 480~GHz, respectively. The resonance frequencies of another SRR change form 370~GHz to 600~GHz, which are presented by the open circles. The dashed lines are the results derived by using three coupled harmonic oscillators model. The coupling strength between each bright mode and the cavity photon is assumed as 84~GHz.}
\end{figure*}

If a MM with two SRRs, which have the same orientation in the unit cell, is placed into the PC cavity, and the MM is excited with a polarization along the gapped sides of the SRRs, then the four-mode coupling phenomenon disappears. This is shown in Fig.~\ref{fig:bright}, which displays simulated transmittance spectra for the case of two bright SRRs per unit cell, as indicated in the inset of Fig.~\ref{fig:bright}a. Both SRRs have their sides with the gap oriented in x-direction, which is also the direction of polarization of the incoming terahertz wave and the direction of the detected electric field. The red and blue curves in Fig.~\ref{fig:bright}a display spectra for SRRs of different size (red curve), respectively the same size (blue curve). The separation distance of the SRRS is in both cases the same, with a value of s = 2~$\mu$m.  
The red curve was obtained for resonance frequencies of the two bright SRRs of 468 and 537~GHz. One observes only three polariton modes. If the resonance frequencies of the SRRs are degenerate (both at 468~GHz), then one even observes only two polariton modes, as shown by the blue curve. 

For more insight into the coupling between the MM with two bright SRRs in the unit cell and the cavity, we show in Fig.~\ref{fig:bright}b a color plot of transmittance spectra as a function of the resonance frequency of one of the SRRs. The resonance frequencies of the other one and of the cavity were kept fixed at 468 and 480~GHz, respectively. Also the distance between the two bright SRRs was kept fixed at 2~$\mu$m. The graph shows an anti-crossing behavior of two modes, an upper (UP) and a lower (LP) polariton mode, with an additional third polariton mode (MP) between them. Approaching the waist region of the anti-crossing, where the resonances of the two SRRs become degenerate, from either above or below the waist, the third mode looses oscillator strength and vanishes at the degeneracy point. A fourth polariton mode does not appear, which is evidence that the second cavity mode is not excited and hence does not take part in the coupling process. The interaction is limited to the two MM eigenmodes and a single cavity mode. We can fit the UP, middle polariton (MP) and LP with the coupled three harmonic oscillators model\cite{Meng2021}. The dash lines derived from the coupled three harmonic oscillators model nicely fit the three polariton branches of the simulation. The open circles represent the varying resonance frequency of the bare SRR, derived from CST simulation. It is interesting to note that the MP diminishes as the resonances of SRRs superimpose with each other, which also represents a polariton dark mode\cite{Zhang2015}. 

\section*{References}

\bibliography{cavity-lib}




\end{document}